\documentclass[12pt]{article}
\usepackage{amsmath}
\usepackage{graphicx,psfrag,epsf}
\usepackage{enumerate}
\usepackage{natbib}
\usepackage{textcomp}
\usepackage[hyphens]{url} 
\usepackage{hyperref}

\newcommand{\blind}{0}

\usepackage{color}
\usepackage{fancyvrb}

\DefineVerbatimEnvironment{Highlighting}{Verbatim}{commandchars=\\\{\}}
\usepackage{framed}
\definecolor{shadecolor}{RGB}{248,248,248}
\newenvironment{Shaded}{\begin{snugshade}}{\end{snugshade}}

\newcommand{\AttributeTok}[1]{\textcolor[rgb]{0.77,0.63,0.00}{#1}}

\newcommand{\ConstantTok}[1]{\textcolor[rgb]{0.00,0.00,0.00}{#1}}

\newcommand{\DecValTok}[1]{\textcolor[rgb]{0.00,0.00,0.81}{#1}}

\newcommand{\FunctionTok}[1]{\textcolor[rgb]{0.00,0.00,0.00}{#1}}

\newcommand{\NormalTok}[1]{#1}

\newcommand{\OtherTok}[1]{\textcolor[rgb]{0.56,0.35,0.01}{#1}}

\newcommand{\SpecialCharTok}[1]{\textcolor[rgb]{0.00,0.00,0.00}{#1}}

\newcommand{\StringTok}[1]{\textcolor[rgb]{0.31,0.60,0.02}{#1}}

\providecommand{\tightlist}{%
  \setlength{\itemsep}{0pt}\setlength{\parskip}{0pt}}

\usepackage{longtable,booktabs,array}
\usepackage{calc} 
\usepackage{etoolbox}
\makeatletter
\patchcmd\longtable{\par}{\if@noskipsec\mbox{}\fi\par}{}{}
\makeatother
\IfFileExists{footnotehyper.sty}{\usepackage{footnotehyper}}{\usepackage{footnote}}
\makesavenoteenv{longtable}

\usepackage{footmisc}
\newcommand{\rfun}[1]{\verb|#1()|}

\usepackage{float}
\usepackage{footmisc}

\begin{document}

\def\spacingset#1{\renewcommand{\baselinestretch}%
{#1}\small\normalsize} \spacingset{1}


\if0\blind
{
  \title{\bf \texttt{rtables} - A Framework For Creating Complex
Structured Reporting Tables Via Multi-Level Faceted Computations}

  \author{
        Gabriel Becker \thanks{Both authors gratefully acknowledge the
following - F. Hoffmann-La Roche Ltd for funding; the NEST team within
Roche for support, testing, and feedback; Jana Stoilova for crucial
design feedback early in the project; Daniel Sabanés Bové for driving
the internal adoption at Roche; and Di Cook, Jayaram Kancherla,
Juha-Pekka Perttola, and Nick Paszty for valuable pre-submission
review.} \\
    No Affiliation\\
     and \\     Adrian Waddell \\
    Genentech Inc.~South San Francisco, USA\\
      }
  \maketitle
} \fi

\if1\blind
{
  \bigskip
  \bigskip
  \bigskip
  \begin{center}
    {\LARGE\bf \texttt{rtables} - A Framework For Creating Complex
Structured Reporting Tables Via Multi-Level Faceted Computations}
  \end{center}
  \medskip
} \fi

\bigskip
\begin{abstract}
Tables form a central component in both exploratory data analysis and
formal reporting procedures across many industries. These tables are
often complex in their conceptual structure and in the computations that
generate their individual cell values. We introduce both a conceptual
framework and a reference implementation for declaring, generating,
rendering and modeling such tables. We place tables within the existing
grammar of graphics paradigm for general statistical visualizations. Our
open source \texttt{rtables} software implementation utilizes these
connections to facilitate an intuitive way to declare complex table
structure and construct those tables from data. In the course of this
work, we relax several constraints present in the traditional grammar of
graphics framing. Finally, \texttt{rtables} models instantiated tables
as tree structures, which allows powerful, semantically meaningful and
self-describing queries and manipulations of tables after creation. We
showcase our framework in practice by creating complex, realistic
example tables.
\end{abstract}

\noindent%
\vfill

\newpage
\spacingset{1.1} 

\spacingset{1.1}

\spacingset{1.1}

\hypertarget{introduction}{%
\section{Introduction}\label{introduction}}

Tabular summaries of information are a key method of both extracting and
communicating information about a dataset or system. Complex tables are
crucial to the analysis and reporting process in many contexts; this
includes the pharmaceutical industry, where clinical trial reporting
tables are an integral part of both the internal decision-making and
external regulatory review processes.\citep{fda_ich_e3, lancet_sub}
Other contexts with formal reporting table requirements include standard
reporting by governmental statistical bureaus \citep{us_bls}, and
financial statements \citep{sec_reporting}. Additionally, reporting
tables are often included in news and academic articles in various
forms.

`Table' is an overloaded term in computing; it can refer to a
rectangular dataset intended primarily for consumption by computers
(e.g., during analysis), or it can refer to a rectangular display of
information or elements intended for human consumption. This conceptual
distinction is crucial, despite some software's willingness to allow
users to create `tables' which mix elements across these purposes
\citep{broman_woo_2018}.

Throughout this work we will use the term \emph{table} exclusively as
shorthand for \emph{reporting table}: a table intended for human
consumption which aggregates and/or summarizes some underlying data. We
will use the term \emph{dataset} to describe a rectangular set of data
-- whether raw or aggregate -- intended for machine consumption.

Numerous software packages, both within and outside of the context of R,
support the creation of tables. Microsoft Excel's Pivot Tables
\citep{excel_citation} and SAS's \texttt{PROC\ TABULATE}
\citep{SAS_citation} declare tabulations and cell value derivations in
terms of variables and functions, following the precedent set by the US
Bureau of Labor Statistics' Table Producing Language (TPL)
\citep{tpl_citation}. Variables are used both to partition the data and
to specify aspects of it to be summarized. Categorical variables define
natural partitions of a dataset, and these tabulation systems use this
to define subsets of data that correspond to rows, columns, and cells.
They also allow nesting, i.e.~consecutively subgrouping the data based
on a sequence of variables.

\citet{tables_rpkg}'s \texttt{tables} R package provides a robust port
of SAS's \texttt{PROC\ TABULATE} API where users generate nicely
rendered tables by specifying the partitioning variables, analysis
variables, and summary functions via formulas \citep{formulas}. R's
\verb|xtabs()| function also provides an interface where users create
tables by declaring the table structure via a formula, though nesting of
partitions and customization of the computation performed on each subset
is not supported. Numerous other packages implement cell value
derivation and table rendering frameworks for specific types of tables
\citep{gtsummary_rpkg, tableone_rpkg}.

The approach of declaring tabulation by categorical variables is also
implemented by R's \verb|aggregate()| \citep{base_rpkg}, which does not
support nesting, as well as \texttt{dplyr}'s \verb|group_by()| into
\verb|summarise()| workflow \citep{dplyr_rpkg} and \texttt{data.table}'s
\texttt{{[},\ by=.{]}} functionality \citep{data.table_rpkg} which do
support nesting. These functions do not generate \emph{tables}, however,
but rather datasets of potential cell values.

Numerous R \citep{base_rpkg} packages provide fine-grained control and
customization of the formatting of tables rendered into HTML
\citep[\citet{tfrmt_rpkg}]{formattable_rpkg, gt_rpkg}, LaTeX and HTML
\citep{xtable_rpkg, huxtable_rpkg} or numerous output formats
\citep{pander_rpkg, flextable_rpkg, ftExtra_rpkg}. These approaches do
not provide mechanisms for \emph{computing} the cell values to be
displayed; rather they typically operate on a pre-existing dataset of
cell values or model fit object and \emph{organize} its elements into a
table. Indications of complex structure, such as the appearance of
multi-level grouping in column or row space are typically achieved by
altering how a table should be \emph{displayed} -- often via
post-processing -- rather than being modeled in the table object.

Beyond R, \citet{mckinney2011pandas}'s Pandas implements \emph{pivot
tables}, which support hierarchical grouping and complex aggregation in
Python. \citet{vink_polars}'s Polars, meanwhile, implements
\texttt{groupby} mechanics similar to \texttt{dplyr}'s in both Python
and Rust.

Clinical trial tables, in particular, can be complex both in
two-dimensional structure and in the data manipulations and statistical
calculations required to derive the individual cell values. These
aspects of tabulation have a causal relationship not often exploited in
existing tabulation frameworks: \emph{table structure implies the
subsetting necessary to perform individual cell-value derivation, and
vice versa}.

Furthermore, the information encoded in this relationship remains useful
beyond cell value derivation. Table rendering (including complex
pagination), table post-processing, and quality-control checking can all
make powerful use of this contextual information if retained.

We have developed a general tabulation framework - informed by the
specific needs of clinical trial reporting - which retains and uses this
intrinsic relationship between table structure, semantic meaning, and
required computations. Our approach differs from and conceptually
extends existing table-creation approaches in a number of ways. Like
SAS, Excel, and the \texttt{tables} package, we allow complex
tabulations to be declared symbolically. We generalize the declaration
of these table structures beyond simple partitioning by categorical
variables. We also fully separate the declaration of reusable, symbolic,
data-instance agnostic \emph{table layouts} from the process of
performing the cell value derivations and constructing the table.
Finally, the resulting tables are represented as a tree structure which
reflects tabulations perform to create them, allowing for semantically
meaningful subsetting and manipulation of the table's elements.

We present here three aspects of our work on complex reporting tables.
First, the \texttt{rtables} software itself and its ability to create
complex tables are showcased in Section \ref{creating-tables}. Secondly,
we place tables within \citet{wilkinson}'s overarching \emph{grammar of
graphics} and \citet{ggplot2_rpkg}'s extension thereof in Section
\ref{DST-as-FST}, and use this to motivate a conceptual framework for
declaring and reasoning about data-summarizing table structures in
Section \ref{layouting}. Finally in Section \ref{table-structure} we
present an object model for instantiated tables which retain semantic
information about the complex aggregations performed during their
creation. We then conclude with a discussion of known limitations and
future directions in Section \ref{discussion}.

\hypertarget{creating-tables}{%
\section{\texorpdfstring{Creating Structured Tables With
\texttt{rtables}}{Creating Structured Tables With }}\label{creating-tables}}

Before discussing the details of \texttt{rtables}' conceptual framework
and design, we first use it to create some illustrative tables and
discuss a case study of production level tables creation for clinical
trial analyses and reporting.

\hypertarget{example-tables-with-code}{%
\subsection{Example Tables With Code}\label{example-tables-with-code}}

Creating a table with \texttt{rtables} is a two-part process. First we
\emph{declare} the conceptual structure of the table by creating a
layout. Secondly, we \emph{apply} that layout to a dataset to construct
a table. The details of layout construction are discussed in Section
\ref{layouting}. For detailed discussions of any of the functions used
in this section, we refer readers to the \texttt{rtables} package
documentation.

The \texttt{rtables} API, is fully general with no assumptions based on
any particular data standards built in. In these examples, we will use
synthetic data\footnote{provided as datasets in \texttt{formatters} and
  originally generated using the open source\texttt{random.cdisc.data} R
  package.} generated to the \texttt{CDISC} ADaM standards
\citep{CDISC_citation} for clinical trial data for our example tables.
Furthermore, we will use the default \texttt{ASCII} text rendering, but
rendering to other formats including \texttt{PDF}, \texttt{HTML} and
\texttt{RTF} is supported.

Our first example table is an adverse events table which summarizes
event occurrences at two hierarchical levels of specificity, seen in
Figure \ref{tbl:ae1}.\footnote{All code used throughout this paper is
  provided in the supplementary materials.}

\spacingset{1.1}

\spacingset{1.1}

\spacingset{1.1}

\spacingset{1.1}

\spacingset{1.1}

\begin{Shaded}
\begin{Highlighting}[]
\NormalTok{lyt }\OtherTok{\textless{}{-}} \FunctionTok{basic\_table}\NormalTok{(}\AttributeTok{show\_colcounts =} \ConstantTok{TRUE}\NormalTok{) }\SpecialCharTok{|\textgreater{}}
  \FunctionTok{split\_cols\_by}\NormalTok{(}\StringTok{"ARM"}\NormalTok{) }\SpecialCharTok{|\textgreater{}}
  \FunctionTok{analyze}\NormalTok{(}\StringTok{"USUBJID"}\NormalTok{, }\AttributeTok{afun =}\NormalTok{ s\_events\_patients) }\SpecialCharTok{|\textgreater{}}
  \FunctionTok{split\_rows\_by}\NormalTok{(}\StringTok{"AEBODSYS"}\NormalTok{, }\AttributeTok{child\_labels =} \StringTok{"visible"}\NormalTok{,}
                \AttributeTok{split\_fun =} \FunctionTok{trim\_levels\_in\_group}\NormalTok{(}\StringTok{"AEDECOD"}\NormalTok{)) }\SpecialCharTok{|\textgreater{}}
  \FunctionTok{summarize\_row\_groups}\NormalTok{(}\StringTok{"USUBJID"}\NormalTok{, }\AttributeTok{cfun =}\NormalTok{ s\_events\_patients) }\SpecialCharTok{|\textgreater{}}
  \FunctionTok{analyze}\NormalTok{(}\StringTok{"AEDECOD"}\NormalTok{, table\_count\_once\_per\_id,}
          \AttributeTok{show\_labels =} \StringTok{"hidden"}\NormalTok{, }\AttributeTok{indent\_mod =} \SpecialCharTok{{-}}\DecValTok{1}\NormalTok{)}

\NormalTok{tbl }\OtherTok{\textless{}{-}} \FunctionTok{build\_table}\NormalTok{(lyt, adae, }\AttributeTok{alt\_counts\_df =}\NormalTok{ adsl, }\AttributeTok{hsep =} \StringTok{"{-}"}\NormalTok{)}
\NormalTok{tbl}
\end{Highlighting}
\end{Shaded}

\begin{figure}[ht!]
\begin{verbatim}
                                           ARM A          ARM B    
                                          (N=146)        (N=154)   
-------------------------------------------------------------------
Patients with at least one event        114 (78.08%)   150 (97.40%)
Total events                                2060           1058    
GASTROINTESTINAL                                                   
  Patients with at least one event      114 (78.08%)   146 (94.81%)
  Total events                              1344           675     
  ABDOMINAL DISCOMFORT                      106             84     
  ABDOMINAL FULLNESS DUE TO GAS             107             98     
  GINGIVAL BLEEDING                          92             73     
  NAUSEA (INTERMITTENT)                     110            109     
MUSCULOSKELETAL AND CONNECTIVE TISSUE                              
  Patients with at least one event      113 (77.40%)   132 (85.71%)
  Total events                              716            383     
  BACK PAIN                                  73             47     
  WEAKNESS                                  111            123     
\end{verbatim}
\spacingset{1.1}
\caption{\label{tbl:ae1}This adverse event table groups events by body system organ class (\texttt{AEBODSYS}) and coded term (\texttt{AEDECOD}) and counts the number of subjects that experience the particular event at least once. Marginal group summaries at all levels display both total number of events and unique number of subjects in the group.}
\end{figure} 
\spacingset{1.1}

Our next example code generates a table (Figure \ref{tbl:bep_all}) which
summarizes overlapping (non-mutually exclusive) groups of observations.
Here, we create a minimal demographics table which displays basic
summary statistics for two demographic variables
(\texttt{SEX}\footnote{We note here that the variable name SEX comes
  directly from the current CDISC standard. The name does not represent
  a claim by us nor our employers about the relationship between
  biological sex and gender identity.} and \texttt{AGE}) separately for
the biomarker evaluable population (BEP) and all patients within each
arm, and for all patients across the study overall.

\spacingset{1.1}

\begin{Shaded}
\begin{Highlighting}[]
\NormalTok{ex\_adsl2}\SpecialCharTok{$}\NormalTok{BEP }\OtherTok{\textless{}{-}} \FunctionTok{sample}\NormalTok{(}\FunctionTok{c}\NormalTok{(}\StringTok{"BEP"}\NormalTok{, }\StringTok{"Non{-}BEP"}\NormalTok{), }\FunctionTok{nrow}\NormalTok{(ex\_adsl2), }\AttributeTok{replace =} \ConstantTok{TRUE}\NormalTok{)}
\NormalTok{combodf }\OtherTok{\textless{}{-}}  \FunctionTok{tribble}\NormalTok{(}
    \SpecialCharTok{\textasciitilde{}}\NormalTok{valname, }\SpecialCharTok{\textasciitilde{}}\NormalTok{label, }\SpecialCharTok{\textasciitilde{}}\NormalTok{levelcombo,         }\SpecialCharTok{\textasciitilde{}}\NormalTok{exargs,}
    \StringTok{"ALL"}\NormalTok{,    }\StringTok{"All"}\NormalTok{,  }\FunctionTok{c}\NormalTok{(}\StringTok{"BEP"}\NormalTok{, }\StringTok{"Non{-}BEP"}\NormalTok{), }\FunctionTok{list}\NormalTok{())}
\NormalTok{combo\_lev\_fun }\OtherTok{\textless{}{-}} \FunctionTok{add\_combo\_levels}\NormalTok{(combodf, }\AttributeTok{keep\_levels =} \FunctionTok{c}\NormalTok{(}\StringTok{"BEP"}\NormalTok{, }\StringTok{"ALL"}\NormalTok{))}

\NormalTok{lyt2 }\OtherTok{\textless{}{-}} \FunctionTok{basic\_table}\NormalTok{(}\AttributeTok{show\_colcounts =} \ConstantTok{TRUE}\NormalTok{) }\SpecialCharTok{|\textgreater{}}
  \FunctionTok{split\_cols\_by}\NormalTok{(}\StringTok{"ARMCD"}\NormalTok{) }\SpecialCharTok{|\textgreater{}}
  \FunctionTok{split\_cols\_by}\NormalTok{(}\StringTok{"BEP"}\NormalTok{, }\AttributeTok{split\_fun =}\NormalTok{ combo\_lev\_fun) }\SpecialCharTok{|\textgreater{}}
  \FunctionTok{add\_overall\_col}\NormalTok{(}\StringTok{"Overall"}\NormalTok{) }\SpecialCharTok{|\textgreater{}}
  \FunctionTok{analyze}\NormalTok{(}\StringTok{"SEX"}\NormalTok{, }\AttributeTok{afun =}\NormalTok{ counts\_wpcts) }\SpecialCharTok{|\textgreater{}}
  \FunctionTok{analyze}\NormalTok{(}\StringTok{"AGE"}\NormalTok{, }\AttributeTok{afun =}\NormalTok{ two\_val\_summary)}

\FunctionTok{build\_table}\NormalTok{(lyt2, ex\_adsl2, }\AttributeTok{hsep =} \StringTok{"{-}"}\NormalTok{)}
\end{Highlighting}
\end{Shaded}

\begin{figure}[ht!]
\begin{verbatim}
                  ARM A                     ARM B                       
            BEP          All          BEP          All         Overall  
           (N=41)       (N=96)       (N=48)       (N=94)       (N=190)  
------------------------------------------------------------------------
SEX                                                                     
  F      27 (65.9%)   59 (61.5%)   28 (58.3%)   52 (55.3%)   111 (58.4%)
  M      14 (34.1%)   37 (38.5%)   20 (41.7%)   42 (44.7%)   79 (41.6%) 
AGE                                                                     
  Mean      31.9         33.1         36.6         36.2         34.6    
  sd        5.2          6.0          9.0          8.3           7.4    
\end{verbatim}
\spacingset{1.1}
\caption{\label{tbl:bep_all}A table comparing two overlapping groups (BEP vs all patients) seperately within each arm, and collectively against the full patient population for the study.}
\end{figure} 
\spacingset{1.1}

Our next code constructs a table where different columns represent
tabulation of entirely different variables, seen in Figure
\ref{tbl:diffvar}. We use \verb|split_cols_by_multivar()| to define that
type of column structure, and \verb|analyze_colvars()| to specify the
analysis.

\spacingset{1.1}

\begin{Shaded}
\begin{Highlighting}[]
\NormalTok{lyt3 }\OtherTok{\textless{}{-}} \FunctionTok{basic\_table}\NormalTok{() }\SpecialCharTok{|\textgreater{}}
    \FunctionTok{split\_cols\_by}\NormalTok{(}\StringTok{"ARMCD"}\NormalTok{) }\SpecialCharTok{|\textgreater{}}
    \FunctionTok{split\_cols\_by\_multivar}\NormalTok{(}\FunctionTok{c}\NormalTok{(}\StringTok{"AGE"}\NormalTok{, }\StringTok{"BMRKR1"}\NormalTok{)) }\SpecialCharTok{|\textgreater{}}
    \FunctionTok{split\_rows\_by}\NormalTok{(}\StringTok{"SEX"}\NormalTok{) }\SpecialCharTok{|\textgreater{}}
    \FunctionTok{analyze\_colvars}\NormalTok{(three\_val\_summary)}

\FunctionTok{build\_table}\NormalTok{(lyt3, ex\_adsl2, }\AttributeTok{hsep =} \StringTok{"{-}"}\NormalTok{)}
\end{Highlighting}
\end{Shaded}

\begin{figure}[ht!]
\begin{verbatim}
                       ARM A                      ARM B          
                  AGE         BMRKR1         AGE         BMRKR1  
-----------------------------------------------------------------
F                                                                
  Mean           32.1          5.7          34.7          5.5    
  sd              5.9          3.2           7.9          3.3    
  Min - Max   23.0 - 47.0   0.5 - 15.1   23.0 - 58.0   0.6 - 13.9
M                                                                
  Mean           34.6          6.5          38.0          5.7    
  sd              6.1          4.1           8.5          3.2    
  Min - Max   23.0 - 48.0   0.4 - 17.7   23.0 - 62.0   1.8 - 14.2
\end{verbatim}
\spacingset{1.1}
\caption{\label{tbl:diffvar}A table which analyzes two different variables (age, biomarker-1 measurement) as columns, within larger column faceting (by trial arm) and row faceting (sex).}
\end{figure} 
\spacingset{1.1}

Our last code example here demonstrates comparisons against a structural
reference group resulting in the table shown in Figure
\ref{tbl:refgroup}. In particular, this table analyzes a (boolean) drug
response including the fitting of statistical models where
\texttt{"ARM\ A"} is considered the reference.

\spacingset{1.1}

\spacingset{1.1}

\spacingset{1.1}

\begin{Shaded}
\begin{Highlighting}[]
\NormalTok{lyt4 }\OtherTok{\textless{}{-}} \FunctionTok{basic\_table}\NormalTok{(}\AttributeTok{show\_colcounts =} \ConstantTok{TRUE}\NormalTok{) }\SpecialCharTok{|\textgreater{}}
  \FunctionTok{split\_cols\_by}\NormalTok{(}\StringTok{"ARMCD"}\NormalTok{, }\AttributeTok{ref\_group =} \StringTok{"ARM A"}\NormalTok{) }\SpecialCharTok{|\textgreater{}}
  \FunctionTok{analyze}\NormalTok{(}\StringTok{"rsp"}\NormalTok{, s\_proportion, }\AttributeTok{show\_labels =} \StringTok{"hidden"}\NormalTok{) }\SpecialCharTok{|\textgreater{}}
  \FunctionTok{analyze}\NormalTok{(}\StringTok{"is\_rsp"}\NormalTok{, s\_unstrat\_resp, }\AttributeTok{show\_labels =} \StringTok{"visible"}\NormalTok{,}
          \AttributeTok{var\_labels =} \StringTok{"Response Analysis"}\NormalTok{)}

\FunctionTok{build\_table}\NormalTok{(lyt4, ADRS\_BESRSPI, }\AttributeTok{hsep =} \StringTok{"{-}"}\NormalTok{)}
\end{Highlighting}
\end{Shaded}

\begin{figure}[ht!]
\begin{verbatim}
                             ARM A            ARM B             ARM C     
                            (N=134)          (N=134)           (N=132)    
--------------------------------------------------------------------------
Responders               114.0 (85.1%)    90.0 (67.2%)      120.0 (90.9%) 
Non-Responders           20.0 (14.9%)     44.0 (32.8%)       12.0 (9.1%)  
Response Analysis                                                         
  Diff Resp Rates (%)                         -17.9              5.8      
  95% CI (Wald)                           (-27.9, -7.9)     (-1.9, 13.6)  
  p-value (Chi^2 Test)                       0.0006            0.1436     
  Odds Ratio (95% CI)                    0.4 (0.2 - 0.7)   1.8 (0.8 - 3.8)
\end{verbatim}
\spacingset{1.1}
\caption{\label{tbl:refgroup}A table analyzing response rate across arms of our fictional study. We note the two separate \texttt{analyze} blocks which go into different levels of detail.}
\end{figure} 
\spacingset{1.1}

Having now seen the code to create various complex tables with
\texttt{rtables} we take a step back to place tables within the larger
context of statistical visualization. We will then present our framework
for declaring complex tables and our object model for representing and
interacting with them.

\hypertarget{case-study-tlg-catalog}{%
\subsection{Case Study: TLG-Catalog}\label{case-study-tlg-catalog}}

Our framework is used for production table generation at Roche. This can
be seen in the open source
\href{https://insightsengineering.github.io/tlg-catalog/}{TLG-Catalog}
\citep{nest_tlgc} developed and maintained by subject matter experts
within the NEST team. The TLG-Catalog is a collection of more than 220
production ready table templates for use in clinical trials analyses and
reporting.

The statistical and business specific logic in TLG-Catalog tables is
implemented in the open source \texttt{tern} package \citep{tern_rpkg}.
\texttt{tern}, in turn, wraps and utilizes \texttt{rtables}' layouting
and tabulation frameworks.

\hypertarget{DST-as-FST}{%
\section{Tables As Faceted Data Visualizations}\label{DST-as-FST}}

\citet[ch.~11]{wilkinson} notes that tables \emph{are} graphs (emph.
his); he does not, however, elaborate on this or how tables fit
meaningfully into his grammar. We do that now.

Consider a standard two-way frequency table of whether a car is
automatic against its number of gears using the \texttt{mtcars} dataset
included with R:

\spacingset{1.1}

\begin{Shaded}
\begin{Highlighting}[]
\NormalTok{mtcars2 }\OtherTok{\textless{}{-}}\NormalTok{ mtcars}
\NormalTok{mtcars2}\SpecialCharTok{$}\NormalTok{am }\OtherTok{\textless{}{-}} \FunctionTok{factor}\NormalTok{(mtcars2}\SpecialCharTok{$}\NormalTok{am, }\AttributeTok{labels =} \FunctionTok{c}\NormalTok{(}\StringTok{"Man"}\NormalTok{, }\StringTok{"Auto"}\NormalTok{))}
\FunctionTok{table}\NormalTok{(}\AttributeTok{transmission =}\NormalTok{ mtcars2}\SpecialCharTok{$}\NormalTok{am, }\AttributeTok{gears =}\NormalTok{ mtcars2}\SpecialCharTok{$}\NormalTok{gear)}
\end{Highlighting}
\end{Shaded}

\begin{figure}[ht!]
\begin{verbatim}
            gears
transmission  3  4  5
        Man  15  4  0
        Auto  0  8  5
\end{verbatim}
\spacingset{1.1}
\caption{\label{base_r_table}Frequency table of transmission type vs number of gears, built with R's \texttt{table()}.}
\end{figure} 
\spacingset{1.1}

Viewed in the Wilkinsonian paradigm, we can frame the above as a faceted
data visualization. The facets are the rows and columns of the table,
while the subplots each contain the corresponding count rendered as
text, as illustrated by the code in Figure \ref{fig:wilkinson_gg}.

\begin{figure}[h]
\spacingset{1.1}
\begin{verbatim}
DATA: mtcars2
DATA: xpos = constant(.5)
DATA: ypos = constant(.5)
COORD: rect(dim(3,4), dim(1,2))
SCALE: linear(dim(1), min(0), max(1))
SCALE: linear(dim(2), min(0), max(1))
ELEMENT: point(position(xpos*ypos*am*gear), size(0),
               label = summary.count(gear*am))
\end{verbatim}
\caption{\label{fig:wilkinson_gg} Grammar of Graphics code declaring our two-way frequency table.}
\end{figure}

This defines a faceted coordinate system, with 2D frames representing
constant variables (\texttt{xpos} and \texttt{ypos}) embedded within 2D
facet frames representing \texttt{am} and \texttt{gear}. Points are then
``drawn'' -- with size zero -- at the values of the constant variables,
the center of the embedded frames, and labeled with the observation
count.

\begin{figure}[ht]
\includegraphics{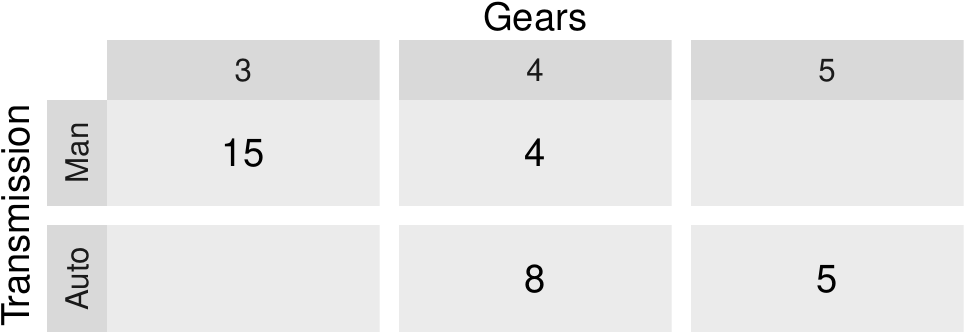} \caption[tbl_in_ggplot]{Our two-way frequency table of transmission type against number of gears, rendered using Wickham's ggplot2 framework.}\label{fig:tbl_in_ggplot}
\end{figure}

The above is equivalent to the following code using
\citet{ggplot2_rpkg}'s layered extension of Wilkinson's grammar for R,
which recapitulates\footnote{Due to implementation details not relevant
  here, \texttt{ggplot2} does not appear to draw geoms when the embedded
  frame for a panel is the empty set, thus leaving blank spaces rather
  than 0s.} our two-way frequency table as a graph in Figure
\ref{fig:tbl_in_ggplot}.

\spacingset{1.1}

\spacingset{1.1}

\spacingset{1.1}

\begin{Shaded}
\begin{Highlighting}[]
\NormalTok{mtcars2[, }\FunctionTok{c}\NormalTok{(}\StringTok{"xpos"}\NormalTok{, }\StringTok{"ypos"}\NormalTok{)] }\OtherTok{\textless{}{-}}\NormalTok{ .}\DecValTok{5}
\FunctionTok{ggplot}\NormalTok{(}\AttributeTok{data =}\NormalTok{ mtcars2,  }\AttributeTok{mapping =} \FunctionTok{aes}\NormalTok{(}\AttributeTok{x =}\NormalTok{ xpos, }\AttributeTok{y =}\NormalTok{ ypos)) }\SpecialCharTok{+}
    \FunctionTok{geom\_text}\NormalTok{(}\AttributeTok{stat =}\NormalTok{ StatCount2) }\SpecialCharTok{+}
\NormalTok{    table\_theme }\SpecialCharTok{+} 
    \FunctionTok{facet\_grid}\NormalTok{(}\AttributeTok{rows =} \FunctionTok{vars}\NormalTok{(am), }\AttributeTok{cols =} \FunctionTok{vars}\NormalTok{(gear), }\AttributeTok{switch =} \StringTok{"y"}\NormalTok{) }\SpecialCharTok{+}
    \FunctionTok{scale\_x\_discrete}\NormalTok{(}\AttributeTok{position =} \StringTok{"top"}\NormalTok{, }\AttributeTok{name =} \StringTok{"Gears"}\NormalTok{) }\SpecialCharTok{+}
    \FunctionTok{scale\_y\_discrete}\NormalTok{(}\AttributeTok{position =} \StringTok{"left"}\NormalTok{, }\AttributeTok{name =} \StringTok{"Transmission"}\NormalTok{)}
\end{Highlighting}
\end{Shaded}

\spacingset{1.1}

\begin{figure}[!ht]
\includegraphics[width=\textwidth]{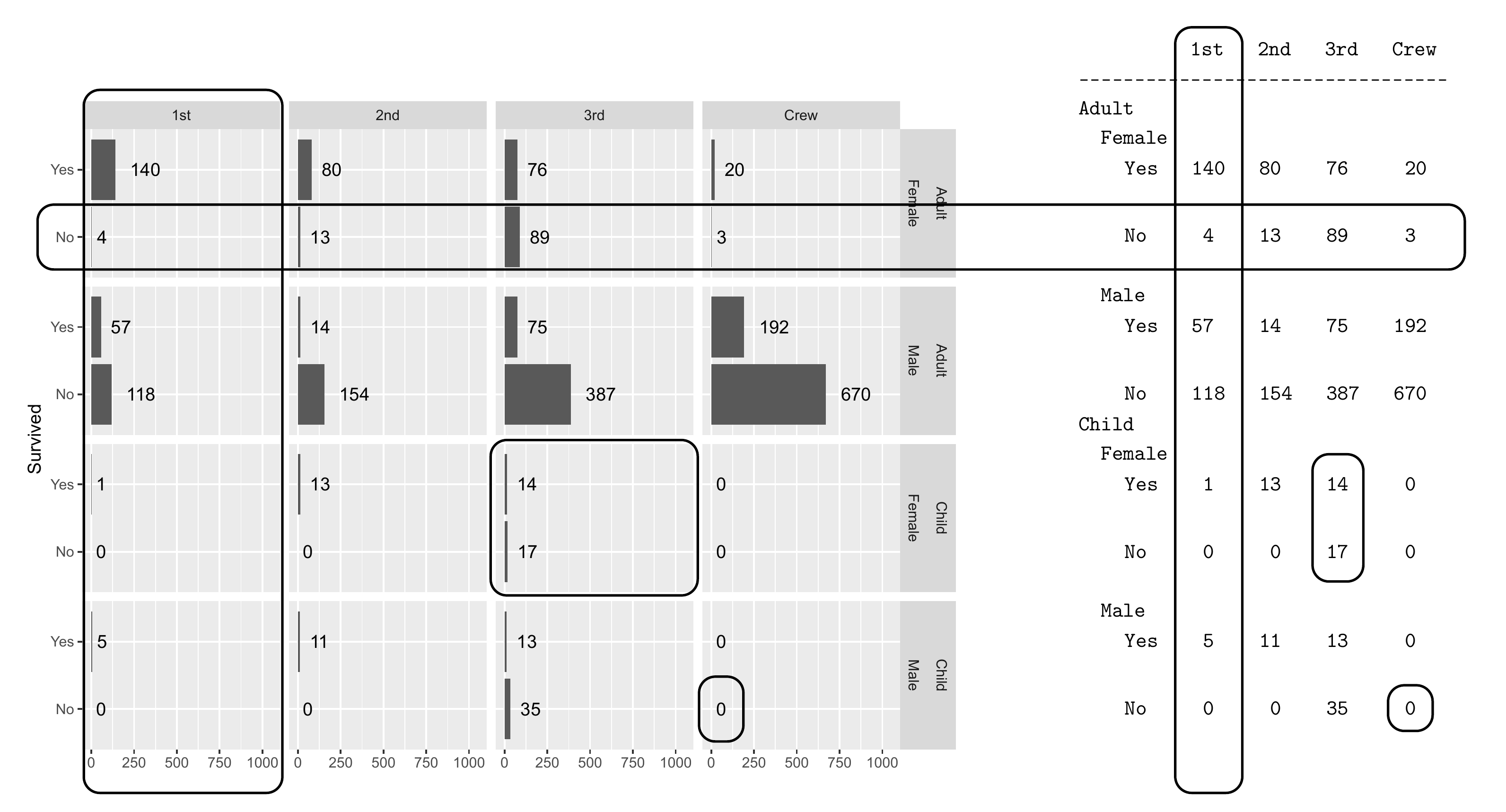}
\spacingset{1.1}
\caption{\label{fig:titanic}Analogous table and faceted barplot summarizing the Titanic death data included with R. Components of the table are highlighted both on the table and the barplot to emphasize their equivalence.}
\end{figure}

Indeed, each conceptual piece of a data-summarizing table maps to an
analogous portion of a corresponding faceted data visualization. We
illustrate this relationship directly in Figure \ref{fig:titanic} and
Table \ref{tab:map_to_grammar}.

Our mapping onto Wilkinson's grammar shows how tables fit within the
existing paradigm while highlighting our generalizations. These
generalizations include 1) relaxing the concept of faceting, 2) allowing
for marginal summarizations at multiple levels in the row-hierarchy, 3)
allowing different facet panes to display different analysis elements.
We motivate and discuss these three extensions in the next section.

\spacingset{1.1}

\begin{longtable}[]{@{}
  >{\raggedright\arraybackslash}p{(\columnwidth - 4\tabcolsep) * \real{0.2262}}
  >{\raggedright\arraybackslash}p{(\columnwidth - 4\tabcolsep) * \real{0.4405}}
  >{\raggedright\arraybackslash}p{(\columnwidth - 4\tabcolsep) * \real{0.3333}}@{}}
\caption{Mapping Table Elements To Plot Components and Grammar Elements
\label{tab:map_to_grammar}}\tabularnewline
\toprule()
\begin{minipage}[b]{\linewidth}\raggedright
Table Component
\end{minipage} & \begin{minipage}[b]{\linewidth}\raggedright
Component in Faceted Barplot
\end{minipage} & \begin{minipage}[b]{\linewidth}\raggedright
In Grammar
\end{minipage} \\
\midrule()
\endfirsthead
\toprule()
\begin{minipage}[b]{\linewidth}\raggedright
Table Component
\end{minipage} & \begin{minipage}[b]{\linewidth}\raggedright
Component in Faceted Barplot
\end{minipage} & \begin{minipage}[b]{\linewidth}\raggedright
In Grammar
\end{minipage} \\
\midrule()
\endhead
Cell & Individual bar & \texttt{ELEMENT} / \texttt{stat\ +\ geom} \\
Row & Multiple bars across subplots (horizontal) & \texttt{ELEMENT} /
\texttt{stat\ +\ geom} \\
Row group & Multiple subplots (horizontal) & \texttt{COORD} /
\texttt{facet\_grid} (Y dim) \\
Group Summary Row & Not Shown - Multiple marginal\footnote{as
  implemented by \citet{ggside_rpkg}'s \texttt{ggside} extension of
  ggplot, though as currently implemented they are limited in the type
  of marginals which can be drawn} subplots & \texttt{COORD} /
\texttt{geom\_xside*} \\
Column or Col. Group & Multiple subplots (vertical) & \texttt{COORD} /
\texttt{facet\_grid} (X dim) \\
Vertical Cell Group & Subplot & \texttt{ELEMENT} /
\texttt{stat\ +\ geom} \\
\bottomrule()
\end{longtable}

\hypertarget{layouting}{%
\section{Pre-Data Layouts - Declaring Table Structure}\label{layouting}}

Faceted plots are made up of (sub)plots which are themselves plotted
within the coordinate system defined by the faceting \citep{wilkinson}.
Furthermore, we have shown that tables \emph{are} faceted data
visualizations. At its core, then, a table has two components:

\begin{enumerate}
\def\labelenumi{\arabic{enumi}.}
\tightlist
\item
  A 2-dimensional faceting structure (\texttt{COORD}), and
\item
  a set of computations to be performed within each facet
  (\texttt{ELEMENT}).
\end{enumerate}

With \texttt{rtables}, users build up a \emph{pre-data table layout}
which symbolically declares both aspects of their desired table and then
apply it to a dataset to create the table.

\hypertarget{incrementally-declaring-facet-structure-for-tables}{%
\subsection{Incrementally Declaring Facet Structure For
Tables}\label{incrementally-declaring-facet-structure-for-tables}}

In \texttt{rtables}, we declare facet structure by repeatedly
\textbf{splitting} a \emph{pre-data table layout} independently in both
row and column dimensions. At each step, we are `splitting' each
existing facet in the relevant dimension further by nesting additional
faceting within the current structure\footnote{as opposed to
  \citet{ggplot2_rpkg}'s implementation of facets in \texttt{ggplot2},
  in which the facet structure is fully declared in a single instruction}.
This provides the intuitive, pipe-friendly workflow for declaring
complex hierarchical faceting that we saw in
\protect\hyperlink{creating-tables}{Section 2}.

In faceted visualizations, every facet corresponds to a specific subset
of the overall data. Faceting, then, is the act of mapping an incoming
parent dataset into the set of datasets corresponding to the facets the
parent will be split into. In the grammar of graphics, this mapping is a
partitioning of the dataset by a categorical variable. We generalize
this concept by allowing faceting to be \emph{any mapping which goes
maps an input dataset to a collection of one or more subsets of that
dataset}; in particular these subsets are not required to be exhaustive
nor are they required to be mutually exclusive. We call this type of
mapping a \emph{split function}. \texttt{rtables} allows users to
customize split functions at the level of individual faceting
instructions. This allows users to declare table structures where
different levels of faceting use different mappings, even within the
same overall faceting dimension. We saw this in practice in column
structure of the table rendered in Figure \ref{tbl:diffvar}.

Non-mutually-exclusive types of faceting that are particularly useful
for tables include those which add an ``all'' category alongside an
otherwise normal partitioning of the dataset, those that compare a
particular category to all observations - as we saw in Figure
\ref{tbl:bep_all}, and those that split data based on the
\emph{cumulative} quantiles of a continuous variable.

Most commonly, however, split functions will define facets by either
partitioning the incoming data based on a categorical variable
(\verb|split_rows_by()|, \verb|split_cols_by()|) or selecting different
individual variables from the full incoming dataset
(\verb|split_cols_by_multivar()|, \verb|split_rows_by_multivar()|).
Partitioning based on categorical variables is equivalent to traditional
faceting implemented in \texttt{ggplot2} in both dimensions, and to
SQL's \texttt{GROUP\ BY} in the row dimension. \citep{ggplot2_rpkg, sql}

Splits in row-space always define structural \emph{groups} of rows;
individual rows are declared by \textbf{summarizing} row groups and
\textbf{analyzing} data within them. We present these aspects of our
framework below.

\hypertarget{declaring-cell-values---summarizing-and-analyzing}{%
\subsection{Declaring Cell Values - Summarizing and
Analyzing}\label{declaring-cell-values---summarizing-and-analyzing}}

The cell contents of a table are declared orthogonally from its faceting
structure, via interspersed use of the \emph{analyze} and
\emph{summarize row group} verbs (\verb|analyze()|,
\verb|analyze_colvars()|, \verb|summarize_row_groups()|). Analyses map
the data of a facet pane to a collection of cells analogous to a single
subplot, which we call a \emph{vertical cell group}. Applied across all
levels of column-faceting, then, each analyze instruction defines a
group of one or more rows in the resulting table.

Group summaries are similar to analyses with the exception that they can
be declared at any point in the hierarchy of the row faceting structure.
They act as marginal analyses, summarizing the data at higher levels of
aggregation and providing semantic context to any analysis and even
other summary rows nested within their facet. We saw this in practice in
the adverse events table in \protect\hyperlink{creating-tables}{Section
2}.

Analyses and group summaries are specified via \emph{analysis functions}
and \emph{summary functions}, respectively. These functions accept data
associated with a facet and return a \emph{vertical cell group}.
Specifying these functions gives users full control over cell value
derivation for their table.

\hypertarget{table-structure}{%
\section{Table Structure}\label{table-structure}}

\begin{figure}[!ht]
\includegraphics[width=\textwidth]{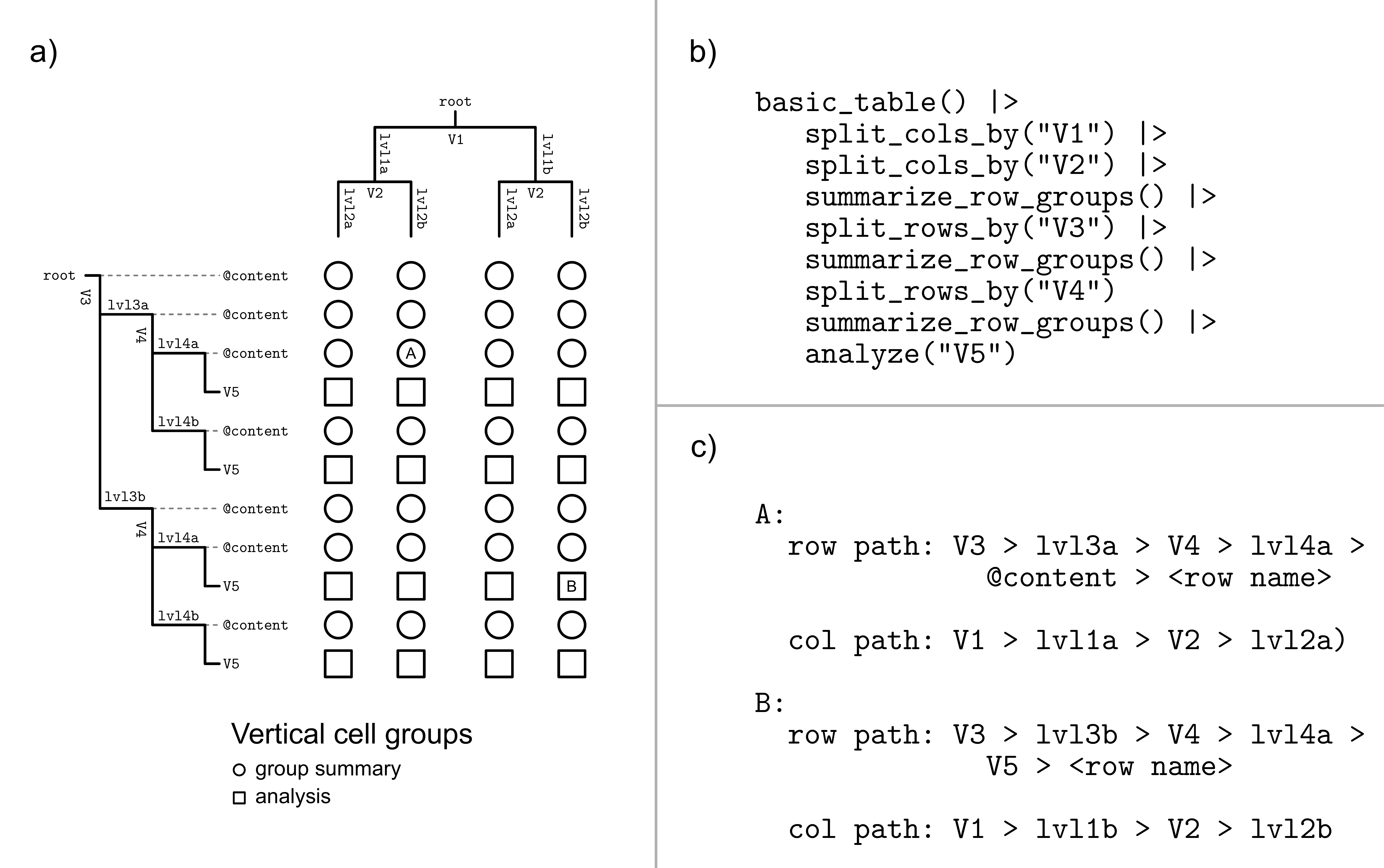}
\spacingset{1.1}
\caption{Relationship between layout, table and paths. a) the
rectangular array of vertical cell groups (VCGs) annotated with the
row and column tree structure; b) code which defines the layout for
the table; c) example row- and column-paths for labelled cells within
the table in (a). Circles in (a) represent VCGs containing marginal
group summaries declared via \texttt{summarize\_row\_groups()} calls in
(b), while squares represent VCGs declared in the \texttt{analyze()}
call. The edge labels within the trees in (a) depict path components
seen in (c).}  \label{fig:structure} \end{figure}

Tables are created by applying a layout to data; \texttt{rtables} models
tables as a set of \emph{analysis} rows hierarchically grouped into
subtables which make up a \emph{TableTree}. Thus our tables are
\emph{row dominant} unlike \texttt{data.frame}s in R which are
\emph{column dominant}. A \texttt{TableTree}'s structure maps directly
to the row-dimension faceting declared by the layout used to create it.
Each faceting instruction maps to a subtable (the split itself)
containing a set of children representing the facets defined by the
split. These children then contain subtables representing either further
nested splitting, or representing the set of rows generated by an
\verb|analyze()| directive. Figure \ref{fig:structure} illustrates this
relationship, which we detail throughout this section.

\spacingset{1.1}

\hypertarget{modeling-context-and-group-summary-information}{%
\subsection{Modeling Context And Group Summary
Information}\label{modeling-context-and-group-summary-information}}

Representing tables as tree structures allows \texttt{rtables} to model
optional \emph{group summary} information which can be displayed when
the table is rendered. These summaries provide visual context for the
individual analysis rows within their facet. We model this contextual
information in two forms: \emph{label rows}, and \emph{content tables}.
Label rows provide context by associating a descriptive label to a
row-space facet. \emph{Content tables} extend the concept of label rows
in two ways: they can contain multiple rows, and they contain non-empty
cell values which provide context within their respective column to the
full facet. We saw both of these types of contextual information in
\protect\hyperlink{creating-tables}{Section 2}.

\emph{Content tables} are created when executing the
\verb|summarize_row_groups()| layouting directive and are a flat
collection of (typically) one or more \emph{content rows}. These content
rows are structurally identical to but semantically different from
analysis rows. They act as an extension -- and when present typically a
replacement -- of label rows by providing both labeling \emph{and} a
marginal summarization of the subtable (and implied group within the
data).

\hypertarget{pathing}{%
\subsection{Pathing}\label{pathing}}

\texttt{TableTree}s' tree structure allows us to define \emph{pathing}
for describing location within a table, similar to but much simpler than
XPath \citep{xpath} defines for XML. We show examples of these paths in
part (c) of Figure \ref{fig:structure}.

Paths provide a semantically meaningful way to describe location within
the structure of a complex table. For example, the paths to the element
labeled \texttt{A} in Figure \ref{fig:structure} correspond to the
elements which represent:

\begin{itemize}
\tightlist
\item
  rows which marginally analyze the data where the values of \texttt{V3}
  and \texttt{V4} are \texttt{lvl3a} and \texttt{lvl4a}, respectively;
  and
\item
  columns within those rows corresponding to the data where the values
  of \texttt{V1} and \texttt{V2} are \texttt{lvl1a} and \texttt{lvl2a},
  respectively.
\end{itemize}

These paths are guaranteed to be descriptive, predictable and
deterministic \footnote{up to non-determinism in, e.g., a custom split
  function}. This allows analysts to use paths \emph{prescriptively},
particularly when the data being summarized adhere to a formal
specification. This is especially powerful, e.g., for selecting cell
values to be included inline in reports such as those in clinical study
reports (CSRs), for defining quality control (cross) checks for a class
of tables, and for defining templates for automatically generated
narrative reports.

Using these paths we can perform semantically meaningful subsetting of a
table in both the row and column dimensions. Pathing also allows us to
perform powerful structurally-aware manipulations of table content after
creation. These notably include modifying or adding: raw cell values;
formatting instructions for table elements; and referential footnotes.

Pathing also facilitates powerful manipulations of a table's structure.
These types of modifications include insertion of rows at
structurally-valid points in the table, as well as sorting or pruning
children based on - typically group summary - cell values within them.
For details on these operations we refer our readers to the
documentation for the \verb|insert_row_at_path()|, \verb|sort_at_path()|
and \verb|prune_table()| functions, respectively.

\hypertarget{analysis-results-datasets-ards}{%
\subsection{Analysis Results Datasets
(ARDs)}\label{analysis-results-datasets-ards}}

The pharmaceutical industry is currently exploring the concept of
Analysis Results Datasets (ARDs)\citep{cdisc_ard} which represents cell
content values as a dataset without the two dimensional table
organization. Motivations for working with ARDs include broad
accessibility and reusability of cell content values.

The concept of ARDs, however, does not address the primary difficulty in
generating complex, multi-faceted tables: the cell value generation
itself. \texttt{rtables} solves this difficulty as demonstrated
throughout this paper. We provide the proof-of-concept \verb|as_ard()|
function which creates a semi-long form \texttt{data.frame} which
combines the cell values from a table with metadata reflecting location
in the faceting structure for that cell. As the ARD specifications
evolves we update our function to meet them.

\hypertarget{discussion}{%
\section{Discussion}\label{discussion}}

We presented the \texttt{rtables} framework and showed how it can be
used to create complex, data summarizing tables including those used for
regulatory submission within the pharmaceutical industry. We then placed
tables within the larger context of data visualizations by showing how
they fit within Wilkinson's grammar of graphics. Finally, we described
an object model for representing and computing on complex tables.

In the next section, we will address some known limitations of both our
conceptual framework and our implementation of it in \texttt{rtables}.
Finally, we will briefly discuss possible future directions for our
work.

\hypertarget{limitations}{%
\subsection{Limitations}\label{limitations}}

Arguably the largest limitation of the \texttt{rtables} approach is,
simply put, that the resulting tables are not \texttt{data.frame}s.
While this is intentional, and powers the computations mentioned in
\href{pathing}{Section 5.2} and others omitted for brevity (e.g.~context
preserving pagination), the fact remains that in some cases, users are
likely to want to perform further analyses on a table's cell values.
Most analysis and visualization tooling in R assume data in a vector or
\texttt{data.frame} format. Our \verb|as_ard()| function derives a
dataset which fully mitigates\footnote{either immediately or after a
  series of standard \texttt{data.frame} transformation steps, depending
  on requirements} such concerns.

Another limitation of our framework is that neither layouts nor the
resulting tables are transposable due to the fact that analyses are
declared exclusively in row-structure space. While this does cause
difficulty in creating certain column-dominant table structures in our
framework, the \verb|split_cols_by_multivar()| layout directive can
force many of these tables into being. Furthermore, in the very few
cases we have seen in practice where the desired gross-structure of a
table was not achievable with \texttt{rtables}, an equivalently
informative row-dominant table that was amenable to our framework has
always existed.

Finally, \texttt{rtables} is complex, and likely to have a rather steep
learning curve for novice users. We do not contest this assertion, but
rather argue -- we hope persuasively -- that this complexity leads
directly to the ease of performing very complex, powerful tabulations
once the user \textbf{is} familiar with it.

\hypertarget{future-directions}{%
\subsection{Future Directions}\label{future-directions}}

Future directions of our work on \texttt{rtables} involve three
components: extensions and advancements to the conceptual framework
underpinning the software, improvements to the functionality of the
software itself, and the reseeding of extensions and generalizations we
made in the table context to the larger arena of visualization. We will
briefly discuss possible future work of each type.

One avenue of future work is the incorporation of statistical plots
\emph{into} tables. While cell values can be anything, and thus -- in
the case of \texttt{grid} \citep{base_rpkg} or \texttt{ggplot2} at least
-- can be graphics themselves, the \texttt{rtables} rendering machinery
does not meaningfully support this. Similarly, support for rendering a
table within or alongside a compound visualization is another piece of
functionality we would like to add in the future.

We also plan to look into the creation of tables where the input data is
not a single rectangular dataset. We expect that because all relational
databases can be denormalized into single (very inefficient) rectangular
datasets, the majority of things supported in \texttt{rtables} should
translate fairly easily into multi-dataset input data, but the work to
prove that -- and to overcome the obstacles surely hiding in the details
-- remain to be done.

Another avenue of future work is making \texttt{rtables}-generated
tables transposable. This could be done by fundamentally modifying our
conceptual model or by implementing (pseudo-)transposition \emph{during
rendering}.

Additionally, \texttt{rtables} has been conceived as a way to create
static tables. Determining where our framework fits within a space where
static tables are still crucial but interactive tabulation is
increasingly important could well define the next stage of
\texttt{rtables}' life-cycle.

Finally, we relaxed a number of restrictions inherent in the grammar of
graphics as it was defined by Wilkinson and extended by Wickham. Chief
amongst these were that:

\begin{enumerate}
\def\labelenumi{\alph{enumi}.}
\tightlist
\item
  Facets need not be a partition of incoming data, but rather a
  generalized grouping;
\item
  Marginal sub-`plots' can be defined at any point in the row faceting
  hierarchy; and
\item
  Elements defining subplots need not be the same across all parts of
  the facet grid.
\end{enumerate}

While these advances were specifically necessary for tables, we feel
they have value in the larger context of visualization generally. We
hope to explore this in future work beyond \texttt{rtables}.

\hypertarget{availability}{%
\section{Availability}\label{availability}}

The \texttt{rtables} R package is available under the commercially
permissive Apache 2.0 open source software license. The current
production version can be installed from CRAN while development versions
can be found -- and issues filed -- at
\url{http://github.com/insightsengineering/rtables}. \texttt{rtables} is
copyright F. Hoffman-La Roche, Ltd.

\bibliographystyle{jasa3}
\bibliography{allrefs.bib}

\end{document}